\documentclass{sf2a-conf}
\usepackage{graphicx}

\def\vsini{$V\!\sin i$}
\def\teff{$T_{\rm eff}$}

\def\logg{$\log~g$}

\def\omc{$\Omega/\Omega_{\rm{c}}$}

\def\ttms{$\frac{\tau}{\tau_{MS}}$}

\begin{document}
\TitreGlobal{SF2A 2007}
\title{On the behaviour of B and Be stars at low metallicity.}


\author{C. Martayan $^{1,2}$} 
\author{Y. Fr\'emat} \address{Royal Observatory of Belgium, 3 avenue circulaire, 1180 Brussels, Belgium}
\author{A.-M. Hubert} \address{GEPI, Observatoire de Paris, CNRS, Universit\'e Paris Diderot; 5 place Jules Janssen 92195 Meudon Cedex, France}
\author{M. Floquet$^{2}$} 
\author{C. Neiner$^{2}$} 
\author{J. Zorec} \address{Institut d'Astrophysique de Paris (IAP), 98bis boulevard Arago, 75014 Paris, France}
\author{D. Baade} \address{European Organisation for Astronomical Research in the Southern Hemisphere, Karl-Schwarzschild-Str. 2, D-85748 Garching b. Muenchen, Germany}
\author{J. Guti\'errez-Soto$^{2,5}$} 
\author{J. Fabregat} \address{Observatorio Astron\'omico de Valencia, edifici Instituts d'investigaci\'o, Poligon la Coma, 46980 Paterna Valencia, Spain}
\author{M. Mekkas$^{2}$} 

\runningtitle{B and Be stars at low metallicity.}

\setcounter{page}{1}

\index{Martayan C.}
\index{Fr\'emat Y.} 
\index{Hubert A.-M.}
\index{Floquet M.}
\index{Neiner C.} 
\index{Zorec J.}
\index{Baade D.}
\index{Guti\'errez-Soto J.} 
\index{Fabregat J.}
\index{Mekkas M.}

\maketitle

\begin{abstract}
We present new results obtained with the VLT GIRAFFE for a large sample of B and Be stars belonging to the Magellanic Clouds, i.e. at
low metallicity. First, we show the effects of the metallicity of the environment on their rotation (linear, angular, and at the ZAMS).
Second, we present the analysis of the effects of metallicity and evolution on the appearance of Be stars. We also new present results 
about the proportions of Be stars to B stars. 
Third, by cross-correlation with large photometric surveys such as MACHO and OGLE, we report on the detection for
the first time of short-term multi-periodicity in 9 Be stars in the Small Magellanic Cloud, which can be interpreted in terms of
pulsations.

\end{abstract}

\section{Introduction}

The origin of the Be phenomenon has given rise to long debates. 
Whether it is linked to stellar evolution or initial formation 
conditions remains a major issue and finally if the Be phenomenon concerns 
only a fraction of B stars or all the B stars. Thus, finding out  differences in
the physical properties of B and Be stars populations belonging to
environments with different metallicity could provide new clues to the understanding of
the Be phenomenon. 

To investigate the influence of metallicity,  star-formation
conditions, and stellar evolution on the Be phenomenon, we have undertaken an
exhaustive study of B and Be stars belonging to young clusters or to the field of the
Small and Large Magellanic Clouds (SMC and LMC), because these galaxies have a
lower metallicity (respectively Z=0.002 and Z=0.004)  than the Milky Way (MW, Z=0.020). 
Since at low metallicity the stellar winds and mass losses are lower than at high metallicity (see Bouret et al. 2003), 
thus the angular mometum loss should be lower, 
and consequently, the rotational velocities are expected to be higher in the SMC and LMC than in the MW (see Meynet \& Maeder 2000).

\section{Observations}

To answer the questions about the metallicity effects, we studied statistically representative samples of B and Be star populations
in environments of different metallicity. The Small and Large Magellanic Clouds are the best tools due to their low metallicity.
We used the VLT GIRAFFE multi-objects spectrograph in MEDUSA mode in the Guaranteed Time Observation programmes of the Observatoire de Paris.
We first prepared the accurate catalogue of possible targets (coordinates better than 0.3", magnitudes) and we preselected the sources
using (V, B-V). Finally, we observed 6.3\% of the B-type stars observable in the SMC GIRAFFE field and 14.3\% in the LMC GIRAFFE field,
which correspond to 346 stars (131 Be, 202 B stars) in the SMC and 177 stars (47 Be, 121 B) in the LMC.

\section{Fundamental parameters}

We make use of the GIRFIT least-squares procedure
(Fr\'emat et al. 2006) to derive the fundamental parameters: effective
temperature (\teff), surface gravity (\logg), projected rotational velocity
(\vsini), and radial velocity (RV). This procedure fits the observations with
theoretical spectra interpolated in a grid of stellar fluxes at the metallicity of the LMC or SMC, computed with the
SYNSPEC programme and from model atmospheres calculated with TLUSTY (Hubeny \&
Lanz 1995, see references therein) or/and with ATLAS9 (Kurucz 1993; Castelli et
al. 1997). 
To derive the luminosity, mass, radius, and ages of stars from
their fundamental parameters, we interpolated in the evolutionary tracks of Schaller
et al. (1992), Charbonnel et al. (1993) calculated for the SMC and LMC metallicities 
and for stars without rotation.

\section{Results}

To study statistically the effects of the metallicity on the rotational velocities, we have to constrain the number of freedom
degrees. Here we must use sub-samples with similar ages, similar masses, and different metallicities by categories: B or Be stars.
The complete results of this section can be retrieved in Martayan et al. (2006, 2007a).

\subsection{Linear rotational velocities}

Merging the stars by mass-ranges, by age-ranges and by types, we obtain the results for the mean rotational velocities in 
Table~\ref{vsiniBmasses} for B stars and in Table~\ref{vsiniBemasses} for Be stars. For comparison, we use the results from Chauville
et al. (2001) and from Zorec et al. (2005) in the Milky Way.
As the distribution of the inclination angle is certainly random, we obtain for each sub-sample the mean rotational velocities at the most
probable angle, thus the equatorial rotational velocities. 

\begin{table*}[th]
\centering
\footnotesize{
\caption{Comparison by mass sub-samples of the mean rotational velocities in the SMC and LMC B stars.
For each sub-sample, the mean age, mean mass, mean \vsini~and the number of stars (N*) are given.
No result is given for massive stars in the SMC, because of their small number.}
\centering
\begin{tabular}{l|cccc|cccc}
\hline
\multicolumn{1}{c}{}&\multicolumn{4}{c}{2-5 M$_{\odot}$}&\multicolumn{4}{c}{5-10 M$_{\odot}$}\\
\hline
	& $<$age$>$ & $<$M/M$_{\odot}$$>$& $<$\vsini$>$ & N* & $<$age$>$ & $<$M/M$_{\odot}$$>$& $<$\vsini$>$ & N* \\
\hline
SMC B stars & 8.0 & 4.0 & 161 $\pm$ 20 & 111 & 8.0 &  6.5 & 155 $\pm$ 17 & 81 \\
LMC B stars & 7.9 & 4.1 & 144 $\pm$ 13 &   6  & 7.6 & 7.2 & 119 $\pm$ 11 & 87 \\
\hline
\end{tabular}
\begin{tabular}{l|cccc|cccc}
\multicolumn{1}{c}{}&\multicolumn{4}{c}{10-12 M$_{\odot}$}&\multicolumn{4}{c}{12-18 M$_{\odot}$}\\
\hline
	& $<$age$>$ & $<$M/M$_{\odot}$$>$ & $<$\vsini$>$ & N*& $<$age$>$ & $<$M/M$_{\odot}$$>$ & $<$\vsini$>$ & N* \\
\hline
SMC B stars &  &  &  &  3 &  &  &  &  2 \\
LMC B stars & 7.3 & 10.7 & 118 $\pm$ 10 &  7  & 6.8 & 13.0 & 112 $\pm$ 10 &  6 \\
\hline
\end{tabular}
\label{vsiniBmasses}
}
\end{table*}

\begin{table*}[!th]
\centering
\footnotesize{
\caption{Comparison by mass sub-samples of the mean rotational velocities for the samples of Be stars in the SMC, LMC and in the MW.
For each sample, the mean age, the mean mass, the mean rotational velocity and the number of stars are given.
Note no low-mass Be star in the LMC.}
\centering
\begin{tabular}{l|cccc|cccc}
\hline
\multicolumn{1}{c}{}&\multicolumn{4}{c}{2-5 M$_{\odot}$}&\multicolumn{4}{c}{5-10 M$_{\odot}$}\\
\hline
	& $<$age$>$ &  $<$M/M$_{\odot}$$>$ & $<$\vsini$>$ & N* & $<$age$>$ & $<$M/M$_{\odot}$$>$ & $<$\vsini$>$ & N* \\
\hline
SMC Be stars & 8.0 &  3.7 & 277 $\pm$ 34 & 14 & 7.6 & 7.6 & 297 $\pm$ 25 & 81 \\
LMC Be stars &  & & &  0 & 7.5 & 7.7 & 285 $\pm$ 20 & 21 \\
MW Be stars & 8.1 & 4.4 & 241 $\pm$ 11 & 18 & 7.6 & 7.3 & 234 $\pm$ 14 & 52 \\ 
\hline
\end{tabular}

\begin{tabular}{l|cccc|cccc}
\multicolumn{1}{c}{}&\multicolumn{4}{c}{10-12 M$_{\odot}$}&\multicolumn{4}{c}{12-18 M$_{\odot}$}\\
\hline
	& $<$age$>$ & $<$M/M$_{\odot}$$>$ & $<$\vsini$>$ & N*& $<$age$>$ & $<$M/M$_{\odot}$$>$ & $<$\vsini$>$ & N* \\
\hline
SMC Be stars & 7.4 &  10.8 &  335 $\pm$ 20 &  13 & 7.2 & 13.5 & 336 $\pm$ 40 &  14\\
LMC Be stars & 7.3 & 11 & 259 $\pm$ 20 & 13 & 7.2 & 14.6 & 224 $\pm$ 30 & 10\\
MW Be stars & 7.2 & 10.6 & 231 $\pm$ 16 & 9 &  6.8 & 14.9 & 278 $\pm$ 10 &  17\\
\hline
\end{tabular}

\label{vsiniBemasses}
}
\end{table*}

These results show clearly that:\\
i) Obviously Be stars rotate faster than B stars whatever the metallicity is.\\
ii) The lower the metallicity is, the higher the rotational velocities are, for B and Be stars 
(see for examples the categories 10-12 M$_{\odot}$ and 5-10 M$_{\odot}$). 
This result is also the first observational proof of the theoretical models from Meynet \& Maeder studies.

\subsection{ZAMS rotational velocities}

\begin{figure}[!htbp]
\centering
\includegraphics[width=4cm, height=7cm, angle=-90]{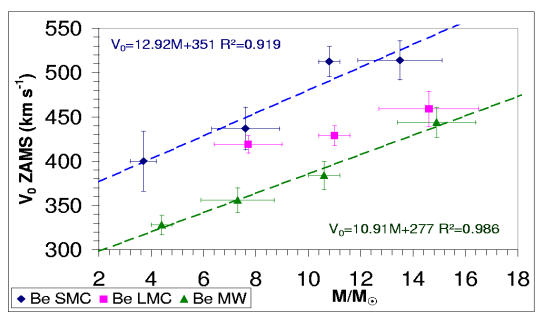}
\caption{ZAMS rotational velocities for the samples of Be stars in the SMC (blue diamonds), in the LMC (pink squares),
and in the MW (green triangles). The dashed lines correspond to the linear regressions. 
Their corresponding equations and correlation coefficient are given in the upper left and lower right corners.
}
\label{V0ZAMSequawind}
\end{figure}

The interpretation of our results requires a set of rotational velocity tracks for masses
between 2 and 20 M$_{\odot}$, for different metallicities corresponding to the MC and the
MW, and for different initial rotational velocities at the ZAMS. We obtain these tracks by
interpolation in the models of the Geneva group. The resulting
distributions of ZAMS rotational velocities for the samples of Be stars in the
SMC, LMC, and MW are shown in Fig.~\ref{V0ZAMSequawind}.

This figure indicates that:\\
i) Whatever the metallicity, the ZAMS rotational velocities of Be stars
depend on their mass.\\
ii) Following the SMC and MW curves, the trend (the gradient) of the ZAMS rotational
velocities of Be stars could be independent of the metallicity.\\
iii) There is an effect of metallicity on the distributions of the ZAMS rotational
velocities: the lower the metallicity, the higher the ZAMS rotational
velocities.\\
iv) There is a limit for the ZAMS rotational velocities below which B stars
will never become Be stars. This limit depends on the metallicity. 
Thus only a fraction of B stars will become Be stars (see also Fig.8 in Martayan et al. 2007a).\\

The star-formation conditions such as the duration of the accretion disk, the accretion rate, the presence of a magnetic field 
could be at the origin of the different ZAMS rotational velocities at different metallicities.

\subsection{Angular rotational velocities}

In the same way, we examined the ratio of angular velocities to their critical angular rotational velocities \omc~for the whole sample
of B and Be stars. The results are given in Table~\ref{angularratio}. They show the effects of the metallicity:  at low metallicity, the
stellar radius should be lower than at thigh metallicity. This, combined to the fact that the linear rotational velocities are higher 
at low metallicity, leads to a higher \omc~ratio.

\begin{table*}[!th]
\centering
\footnotesize{
\caption{Comparison by mass sub-samples of the percentage of the mean angular rotational velocities 
to its critical angular rotational velocity for the samples of B and 
Be stars in the SMC, LMC and in the MW.}
\centering
\begin{tabular}{cccc}
\hline\hline
Z & Galaxy & B stars & Be stars \\
\hline
0.020 & Milky Way & 35 \% & 80 \% \\
0.004 & Large Magellanic Cloud & 37 \% & 85 \% \\
0.002 & Small Magellanic Cloud & 58 \% & 95 \% \\
\hline 
\end{tabular}
\label{angularratio}
}
\end{table*}

\section{Stellar evolution and Be stars appearance}

As in Zorec et al. (2005), we derived the evolutionary status for the Be stars in the LMC and SMC samples.
The percentages of Be stars in the first and in the second part of the Main Sequence (MS) are given in Table~\ref{propBe}. 

\begin{table}[th]
\centering
\footnotesize{
\caption{Proportions in SMC, LMC, and MW of Be stars (in \%) in the upper (\ttms$>$ 0.5) and lower (\ttms$\leq$ 0.5) MS 
for masses $>$ 12 M$_{\odot}$ and $\leq$ 12 M$_{\odot}$. MW values come from from Zorec et al. (2005).}
\centering
\begin{tabular}{l|ccc|ccc}
\hline
\multicolumn{1}{c}{}&\multicolumn{3}{c}{M $>$ 12 M$_{\odot}$}&\multicolumn{3}{c}{M $\leq$ 12 M$_{\odot}$}\\
\hline
  & MW & LMC & SMC                 & MW & LMC & SMC \\
\hline
\ttms$>$0.5 & 30 & 100 & 100      & 65 & 77 & 94 \\
\ttms$\leq$0.5 & 70 & 0 & 0       & 35 & 23 & 6 \\
\hline
\end{tabular}
\label{propBe}
}
\end{table}

\begin{figure}[!htbp]
\centering
\includegraphics[width=3.5cm, angle=-90]{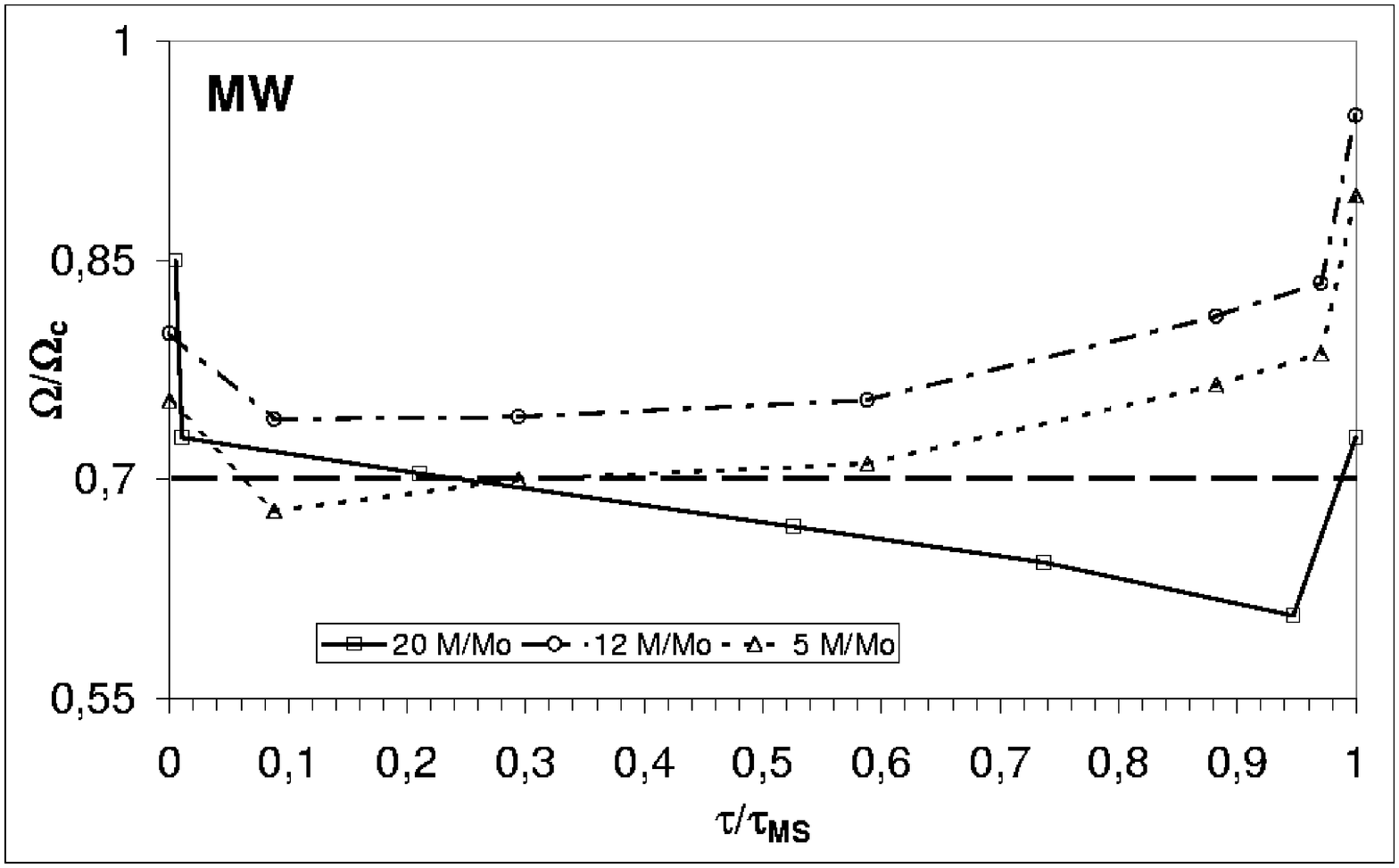}
\includegraphics[width=3.5cm, angle=-90]{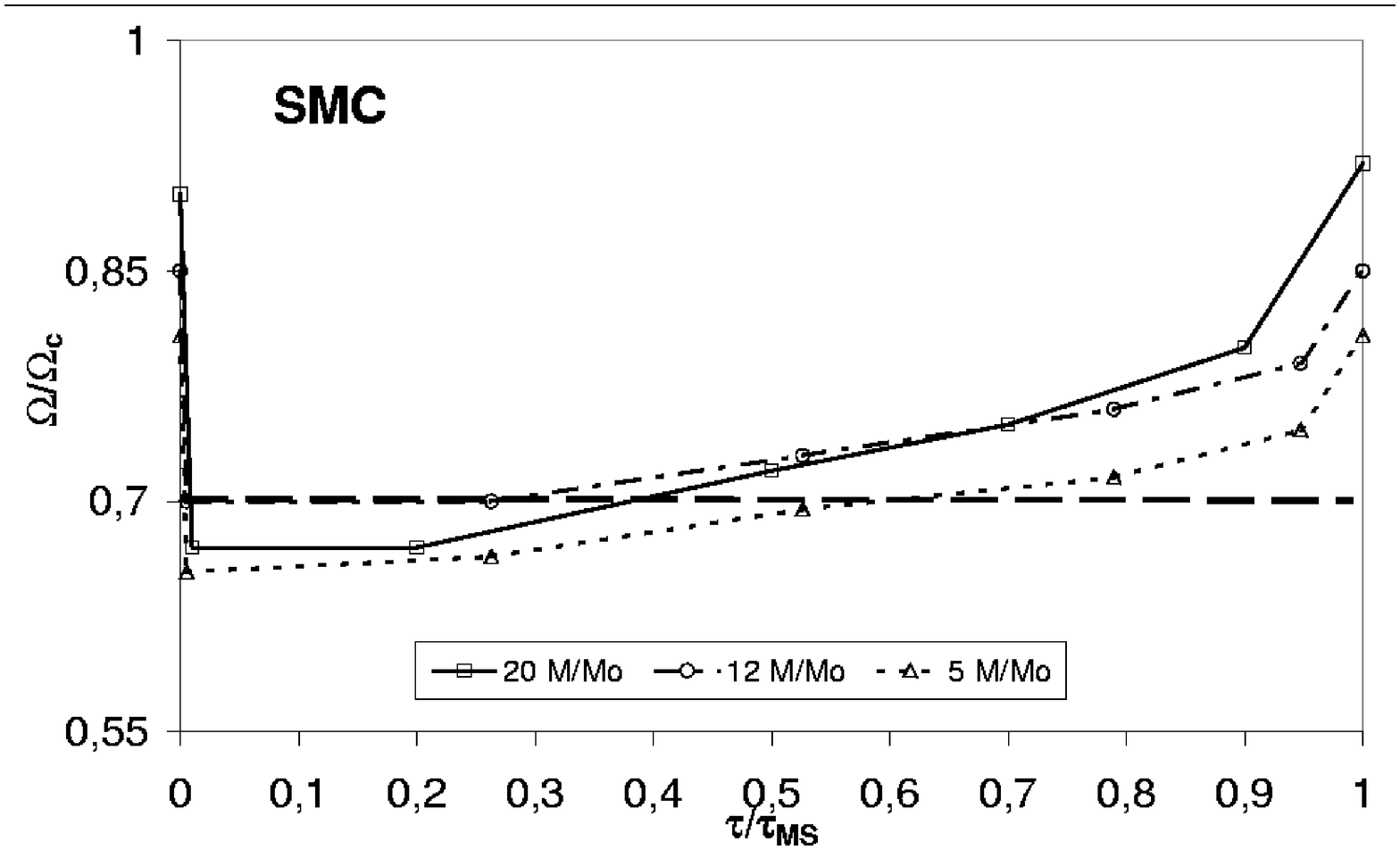}

\caption{Evolution of the \omc~ratio for 3 types of stars:
20 M$_{\odot}$ (squares), 12 M$_{\odot}$ (circles) and 5 M$_{\odot}$ (triangles) 
stars in the MW (top) and in the SMC (bottom).
The stars may become Be stars if \omc$\geq$70\% (noted with a long dashed line).}
\label{evolBeOMmoa}
\end{figure}

These results indicate that the less massive Be stars are mainly in the second part of the MS whatever the metallicity.
The more massive Be stars are mainly in the first part of the MS in the MW at the opposite of those at low metallicity (SMC and LMC), which
are all in the second part of the MS. 
The results could be explained by the evolution of the angular rotational velocities at different metallicity and by the fact that Be stars
appear with at least \omc $\ge$ 70 \%.  Fig.~\ref{evolBeOMmoa} shows the evolution of the \omc~across the time for stars in the MW and
in the SMC and shows the area of existing Be stars.
The complete results of this section can be retrieved in Martayan et al. (2006, 2007a).

\section{Photometric variability and pulsations}

By cross-correlation of the GIRAFFE samples with the MACHO photometric survey, we obtained light-curves for the main part of the stars.
Then by using different methods we searched for photometric variability and short-term periodicity in Be stars.
Finally, we found 13 Be stars with short-term periodicity and 9 of them are multi-periodic pulsators, which plaids in favour of pulsations. 
Moreover, the detected periods fall in the range of SPB-type pulsating modes (from 0.40 to 1.60 days).
The current theoretical models (see Pamyatnykh 1999) do not predict the presence of pulsational instabilities in
massive stars at low metallicities, i.e. in the SMC.
The complete results of this section can be retrieved in Martayan et al. (2007b).


\section{Proportions of Be stars to B stars in the SMC}

As the rotational velocities at low metallicity are higher than at high metallicity, it is probable that the number of Be stars to B stars
increase with decreasing metallicity. Maeder et al. (1999) found a small increase of this ratio with decreasing metallicity but in the SMC
statistics, only one open cluster was used. A large survey of Emission Line Stars and specifically of Be stars is needed to improve the 
determination of the proportions of Be stars in the Magellanic Clouds. In the following sections, we present the survey we performed and the
first results obtained.

\begin{figure}[!ht]
\centering
\includegraphics[height=12cm, width= 6cm, angle=-90]{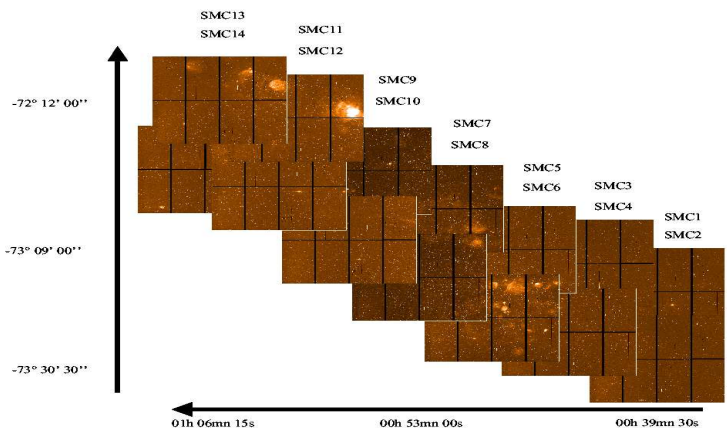}
\caption{The Small Magellanic Cloud as projected on the WFI frames used in this study.}
\label{mapSMC}       
\end{figure}

\subsection{The WFI H$\alpha$ survey of the Magellanic Clouds}
Observations covering much of the Small Magellanic Cloud (see Fig.~\ref{mapSMC}) and the Large Magellanic Cloud
have been obtained in September 2002 with the WFI attached to the 2.2-m MPG Telescope at La Silla.  
The instrument was used in its slitless
spectroscopic mode. To reduce crowding, the length of the spectra was limited by means of a filter with a
bandpass of 7.4 nm centered on H$\alpha$. Unfortunately, a large part of the fields suffers from substantial
non-homogeneous defocusing,  which severely reduces the contrast between stars with and without line emission
at H$\alpha$.
The extraction in 2-D of the spectra was accomplished by means of the
SExtractor software Bertin \& Arnouts (1996).  All in all, about 1 million of the 3 millions of spectra available in the
SMC part of the  survey proved usable.
To recognize and distinguish emission-line stars (ELS) from other objects,  we created the ALBUM package in
IDL. For more information about it, we refer the reader to Martayan et al. (2007c).

\begin{figure}[h!]
\centering
\includegraphics[height=6cm, angle=-90]{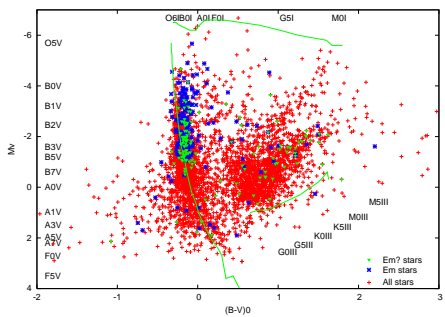}
\caption{(Mv,(B-V)$_{0}$ diagram for the stars cross-correlated in OGLE database. The calibration in spectral
types comes from Lang (1991). Red '+' are for the non-emission line stars, Blue '*' for the emission line
stars, and the green triangles are for the candidate emission line stars.}
\label{fig4}       
\end{figure}

\subsection{Results: proportions}

\begin{figure}[h!]
  \centering
  \includegraphics[width=4cm, angle=-90]{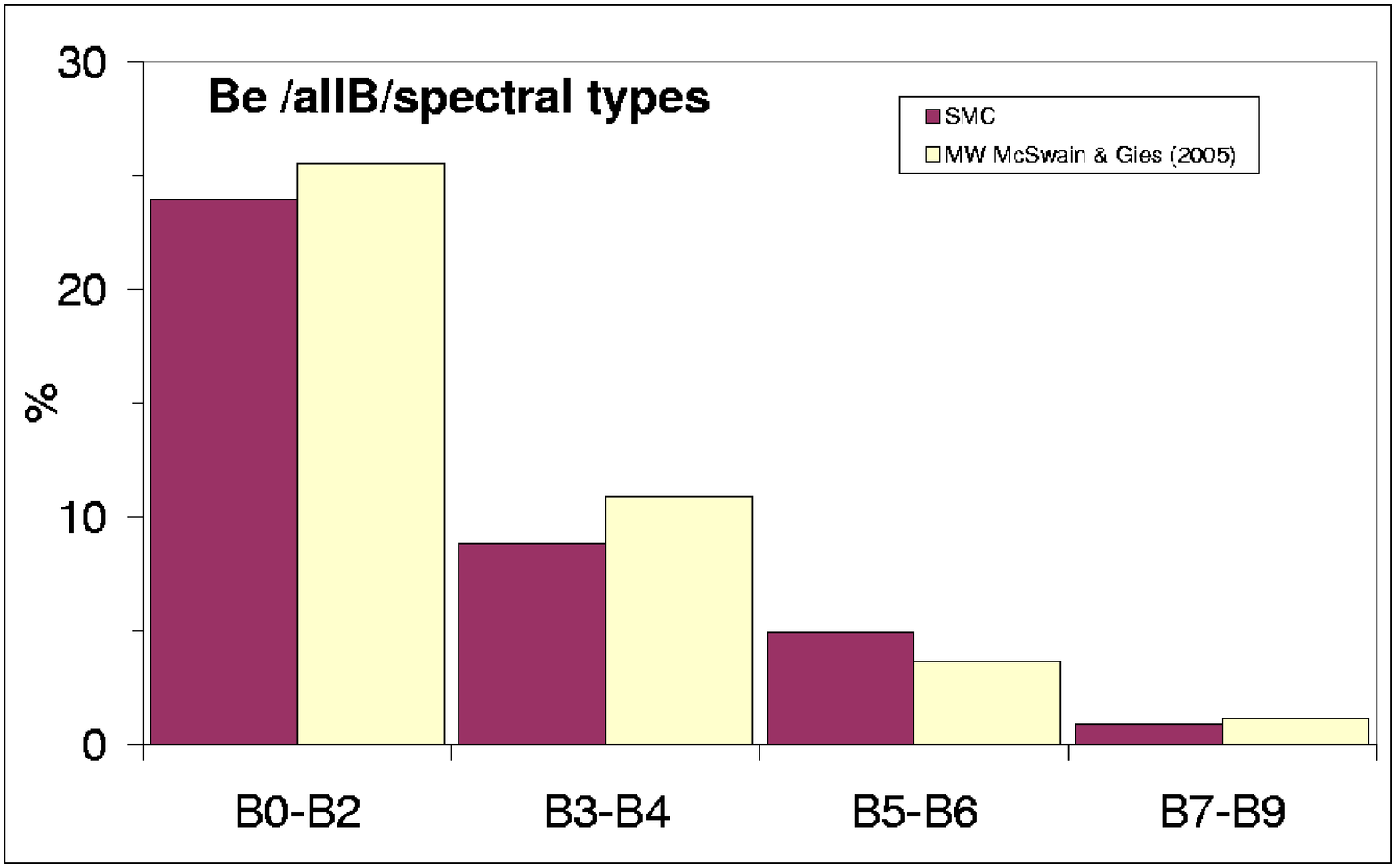}
   \includegraphics[width=4cm, angle=-90]{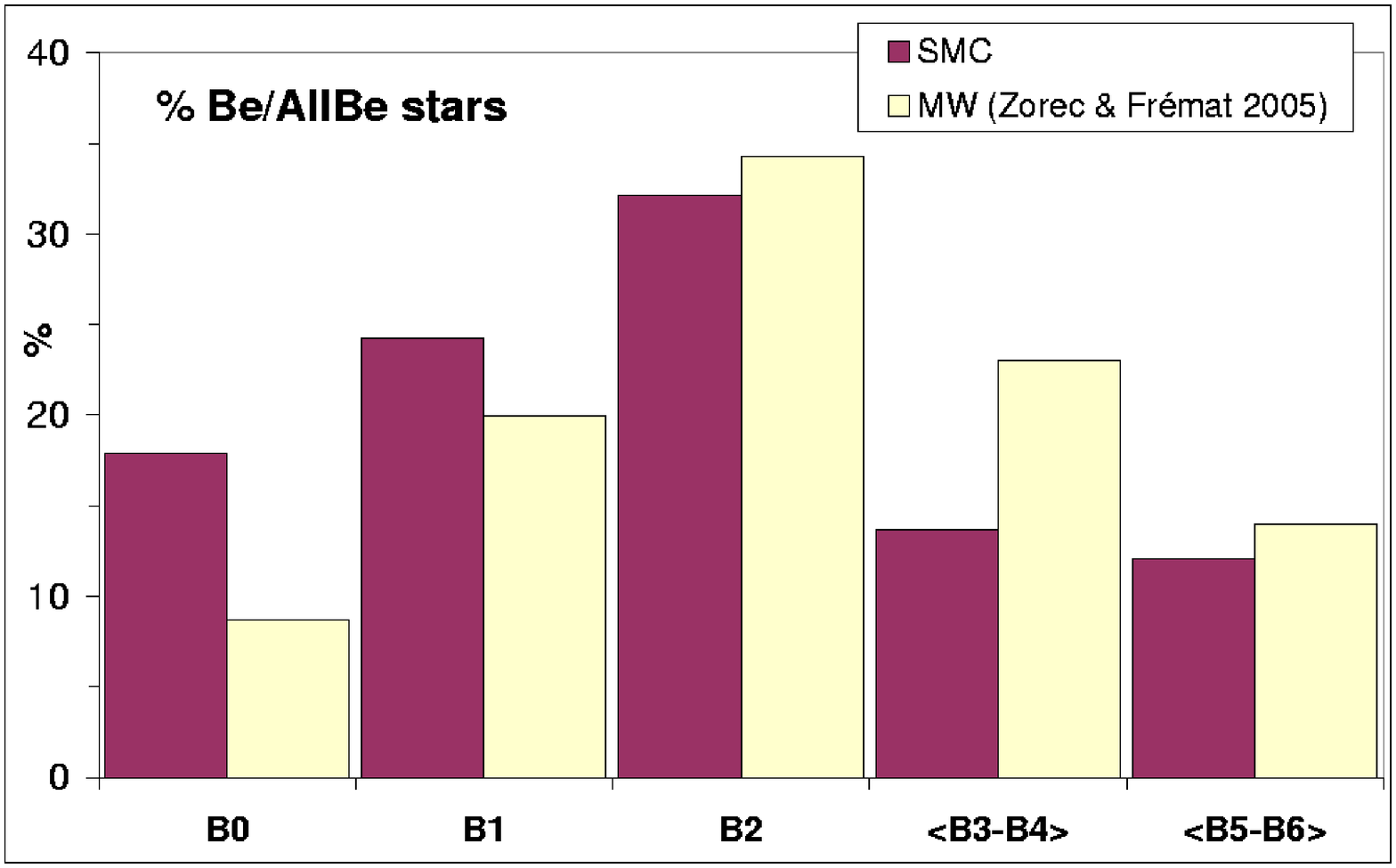}
      \caption{Left: Proportions of Be stars to B stars by spectral types in the SMC and in the Milky Way (from
      McSwain \& Gies 2005).
      Right: Proportions of Be stars to all the Be stars 
      per spectral types in the SMC and in the
      MW (Zorec \& Fr\'emat 2005).
      }
       \label{stats}
   \end{figure}

We have investigated 85 clusters in the SMC with log(age) between 7 and 9 and E[B-V] available from the OGLE survey
For a total of 7741 stars, V, B, and I magnitudes were obtained from the OGLE database (Pietrzy\'nski et al. 1999). 
Fig.~\ref{fig4} displays the combined HR diagram of all clusters with the ELS marked. The results can
be compared in Fig.~\ref{stats}, left,  to the relative  frequencies of Be stars (Be/(B+Be)) in Milky Way clusters (McSwain \& Gies 2005) 
in order to search for any effect of the metallicity on the proportion of Be stars and on the still unknown  reason for the
development of disks around these extremely rapidly rotating stars.  There seems to be a trend in that the
lower the metallicity, the higher the proportion of Be stars is.   This could be explained by higher rotational
velocities in the SMC than in the MW  (Maeder \& Meynet 2001, Martayan et al. 2007b). 
In the SMC as in the MW (Zorec \& Fr\'emat 2005, Be stars in fields) the 
occurrence of Be stars in the 2 galaxies shows a maximum at B2 as shown in Fig.~\ref{stats} right.

\section{Conclusion/perspectives}
We found that: the lower the metallicity is, the higher the linear, angular, and ZAMS rotational velocities are.
We also found that the progenitors of Be stars rotate faster than those of B stars, which indicate that Be stars
correspond to a a sub-group of B-type stars.
The evolution and the appearance of Be stars seem also driven by the metallicity of the environment and by the ZAMS
rotational velocities. At the metallicity of the SMC, we found several multi-periodic pulsators, while pulsations are not foreseen by the theoretical
models. It seems also that there are more Be stars at low metallicity (SMC) than in the MW.
As perspectives of this work, we shall determine the chemical abundances in slow and fast rotators (B and Be stars) at low and high metallicity
and for different stellar evolution degrees. 
At middle term, we shall continue the B-type star populations spectroscopic studies in the Magellanic Clouds and 
in the other local group galaxies with the VLT, the ELT, and with an IR survey at the Dome C.
At long term we shall explore the First Stars (very faint metallicity) of galaxies beyond the local group
with the future multi-objects instrument of the ELT.

\tiny{

}
\end{document}